# Analysis of Big Data Maturity Stage in Hospitality Industry

Neda Shabani*, Arslan Munir[†], and Avishek Bose[‡]

Email: *nshabani@ksu.edu, [†]amunir@ksu.edu, and [‡]abose@ksu.edu

Big data analytics has attracted a lot of attention in recent years from many industries and businesses due to its huge impact on businesses operations, decision making process, and profitability. Some of the industries such as health care heavily rely on big data analytics for their profitability and customer satisfaction. Big data analytics has an extremely significant impact on many areas in all businesses and industries including hospitality. Considering the large number of customers in the hotel industry, the data collected in real-time or near real-time naturally leads to huge *Volumes* of disparate data, arriving at high *Velocity* with considerable *Variety* in format. Furthermore, integration of data collected with different spatio-temporal resolutions introduces *Veracity* and *Variability* to the data. Such properties of the acquired data transform the data analytics challenge into a *Big Data* problem. According to Big-Data Analytics in hotel industry (August 9, 2015), some of these areas are customer segmentation, customer profiling, site selection, forecasting, customer relationship management, menu engineering, productivity indexing, customer associations and sequencing, website optimization, customized marketing, customer value, energy consumption, and investment management. Hospitality industry is one of the industries that receive huge volumes of data. As increasing the revenue and improving guest experience are the core goals of the hospitality industry, having a proper and efficient big data analytics platform can dramatically lead the industry to reach its goals and affect the way the industry runs in all the mentioned areas.

Unfortunately, most of the hotels currently do not have enough knowledge about their acquired data, hence the data remains under-appreciated and under-used asset. For instance, most hotels capture data about customer loyalty, but very few go beyond the data collection stage and actually exploit this data to enhance their knowledge of their guests and develop a better understanding of the behavior of different customer segments, their expectations, and needs (Marr, 2016). However, if hoteliers be able to analyze their captured data properly, they will be able to understand the needs, demands, and expectations of their customer better and will be able to meet those expectations. Also since keeping one guest loyal is much cheaper than bringing one new guest to the hotel, hoteliers have to be aware of their guests' purchase behavior (e.g., frequency, length of stay, time of year) and preferences (e.g., locations, activities, and room types), so that they identify better opportunities to attract the guests, determine the profitable market segmentation, increase the wallet share, and increase the brand loyalty (Big-Data Analytics in hotel industry, 2015).

Red Roof Inn is a hotel chain that increased the business revenue by 10% through the Big Data Analytics (Marr, 2016). As Marr (2016) stated: "The chain's marketing and analytics team worked together to identify openly available public datasets on weather conditions and flight cancellations. Knowing that most of their customers would use web search on mobile devices to search for nearby accommodation, a targeted marketing campaign was launched, aimed at mobile device users in the geographical areas most likely to be affected".

The challenge for hoteliers is to collect, manage, and analyze this big data to provide support in decision making to maximize the hotel profit and improve the guests' satisfaction by offering customized offers and promotions. Currently hotels collect more volumes of data than they can actually manage or analyze, however, hoteliers are aware of the significance of big data and its effectiveness on creating strategic competitive advantage. In order to gain a strategic competitive advantage, hotels need to understand where they are, where they have been, and where they need to go on their big data deployments. This study aims to guide information technology (IT) professionals in hospitality on their big data expedition. In particular, the purpose of this study is to identify the maturity stage of the big data in hospitality industry in an objective way so that hotels be able to understand their progress, and realize what it will take to get to the next stage of big data maturity through the scores they will receive based on the survey.

To identify the maturity stage of big data in hospitality industry, this study uses a survey from Transforming Data With Intelligence (TDWI) Big Data Maturity Model and Assessment Tool. The survey targets the hospitality Information Technology (IT) professionals in the United States who are the member of Hospitality Financial and Technology Professionals (HFTP). The emails of targeted population will be obtained from HFTP and the survey will be sent to them through Qualtrics. The obtained data will be later analyzed based on the TDWI (n.d.) big data maturity stages. According to the TDWI (n.d.), big data has five stages of maturity: *nascent, pre-adoption, early adoption, corporate adoption, and mature/visionary*.

This study will contribute to the public knowledge as well as hospitality IT professionals in order to understand how their big data deployments compare to those of their peers in order to provide advanced insight and support. Also since no scholar has ever researched about the big data maturity stage in hospitality industry, this study fills the gap in the literature review. In addition, this study provides guidance for hotels at the beginning of their big data venture by helping them understand best practices used by other hotels that are more mature in their big data deployments. Hospitality IT professionals can use the result of this study for their big data adoption on organization, infrastructure, data management, governance and analytics level. Finally, this study discusses the implications for increasing the big data maturity and provides directions for future studies. This study doesn't consider the difference between hotel types and will generalize the results for all hotel types, which is a limitation of this study.